\documentclass[prl,aps,epsfig,superscriptaddress,showpacs]{revtex4}
\usepackage{bm}
\usepackage{amsfonts}
\usepackage[dvips]{graphicx}
\usepackage{mathrsfs}
\usepackage[intlimits]{amsmath}
\usepackage[colorlinks, citecolor=red]{hyperref}
\usepackage{color}

\begin{document}

\title{Realization of quantum Maxwell's demon with solid-state spins}
\author{W.-B. Wang$^{1}$, X.-Y. Chang$^{1}$, F. Wang$^{1}$, P.-Y. Hou$^{1}$,
Y.-Y. Huang$^{1}$, W.-G. Zhang$^{1}$, X.-L. Ouyang$^{1}$, X.-Z. Huang$^{1}$,
Z.-Y. Zhang$^{2}$, L. He$^{1}$, L.-M. Duan}
\affiliation{Center for Quantum Information, IIIS, Tsinghua University, Beijing 100084,
PR China}
\affiliation{Department of Physics, University of Michigan, Ann Arbor, Michigan 48109, USA}
\date{\today }

\begin{abstract}
Resolution of the century-long paradox on Maxwell's demon reveals a deep
connection between information theory and thermodynamics. Although initially
introduced as a thought experiment, Maxwell's demon can now be implemented
in several physical systems, leading to intriguing test of
information-thermodynamic relations. Here, we report experimental
realization of a quantum version of Maxwell's demon using solid state spins
where the information acquiring and feedback operations by the demon are
achieved through conditional quantum gates. A\ unique feature of this
implementation is that the demon can start in a quantum superposition state
or in an entangled state with an ancilla observer. Through quantum state
tomography, we measure the entropy in the system, demon, and the ancilla,
showing the influence of coherence and entanglement on the result. A quantum
implementation of Maxwell's demon adds more controllability to this
paradoxical thermal machine and may find applications in quantum
thermodynamics involving microscopic systems.
\end{abstract}

\date{\today }
\maketitle


\section{Introduction}

Maxwell's Demon is a gedanken experiment introduced a century ago by Maxwell
that has a seeming contradiction with the second law of thermodynamics \cite%
{1,2}. Suppose a box full of gas particles is separated into left and right
sides and connected through a small gate controlled by a microscopic being
called the demon. When a gas particle comes towards the gate, the demon only
opens the gate when it observes that the particle is from the left. Due to
this selective opening of the gate, eventually all the gas particles move to
the right side, and the system entropy thus decreases which seemingly
contradicts with the second law of thermodynamics. After decades of
controversies in the interpretation, this paradox is finally resolved by the
observation that the demon has to acquire information about the gas particle
for each gate operation \cite{1,2,3,4,5}. Erasing of this information
transfers a finite amount of work to heat (Landauer's principle) and thus
the demon's own entropy increases which is always no less than the entropy
decrease in the system \cite{1,2,3,4,5,6,7,8}.

The essence of Maxwell's demon experiment is the information acquiring by
the demon and the feedback to the system depending on this information. This
thought experiment has been implemented in several physical systems, where
the entropy decrease and the transferring of heat to work in the system has
been observed \cite{9,9b,10,11,12,13,14,15,16,17}. The demon's operations
could also be implemented by quantum gates using microscopic qubits \cite%
{6,7,8}. A quantum implementation allows the investigation of the role of
quantum coherence and entanglement in the microscopic limit of thermal
machines, a topic that raised significant interest in the rising field of
quantum thermodynamics \cite{18,19,20,21}. In particular, entanglement could
be interpreted as a resource of negative entropy for some thermodynamic
process \cite{20}. Very recently, Maxwell's demon has been demonstrated
using the nuclear magnetic resonance and the superconducting cavity systems
based on quantum gate operations \cite{22,23}. In those experiments, the
demon still starts at a classical state and the role of entanglement has not
been investigated yet.

Here, we report an experimental realization of quantum Maxwell's demon based
on control of solid-state spins which demonstrates the important role of
memory consumption and entanglement in the functionality of a quantum demon.
A key observation to the resolution of Maxwell's demon paradox is that the
operation by the demon consumes its memory and without a memory reset it
will not be able to continuously reduce the system entropy. To demonstrate
this key point, we use electronic spin states to represent the demon's
memory and experimentally observe that the same demon's operation
successively on two nearby nuclear spins reduces the entropy of the first
system but fails to do so for the second one. Then, we prepare the demon in
a quantum superposition state and demonstrate that the entropy decrease in
the system depends on the observation basis. Finally, to demonstrate the
important role of entanglement, we entangle the demon at the beginning with
an ancilla representing an insider observer. For an outside observer without
access to the ancilla's information, the demon's operation fails to reduce
the system entropy. However, for an inside observer with help from the
ancilla, we show that the entropy for both the system and the demon can
decrease while the ancilla's entropy remains the same. This contradiction
with the interpretation of the classical demon therefore experimentally
confirms that the entanglement can serve as a resource of negative entropy
for the quantum Maxwell's demon \cite{20}.

\section{Results}

\subsection{Implementation of quantum Maxwell's demon with solid state spins}

\begin{figure}[tbp]
\centering
\includegraphics [scale = 0.7]{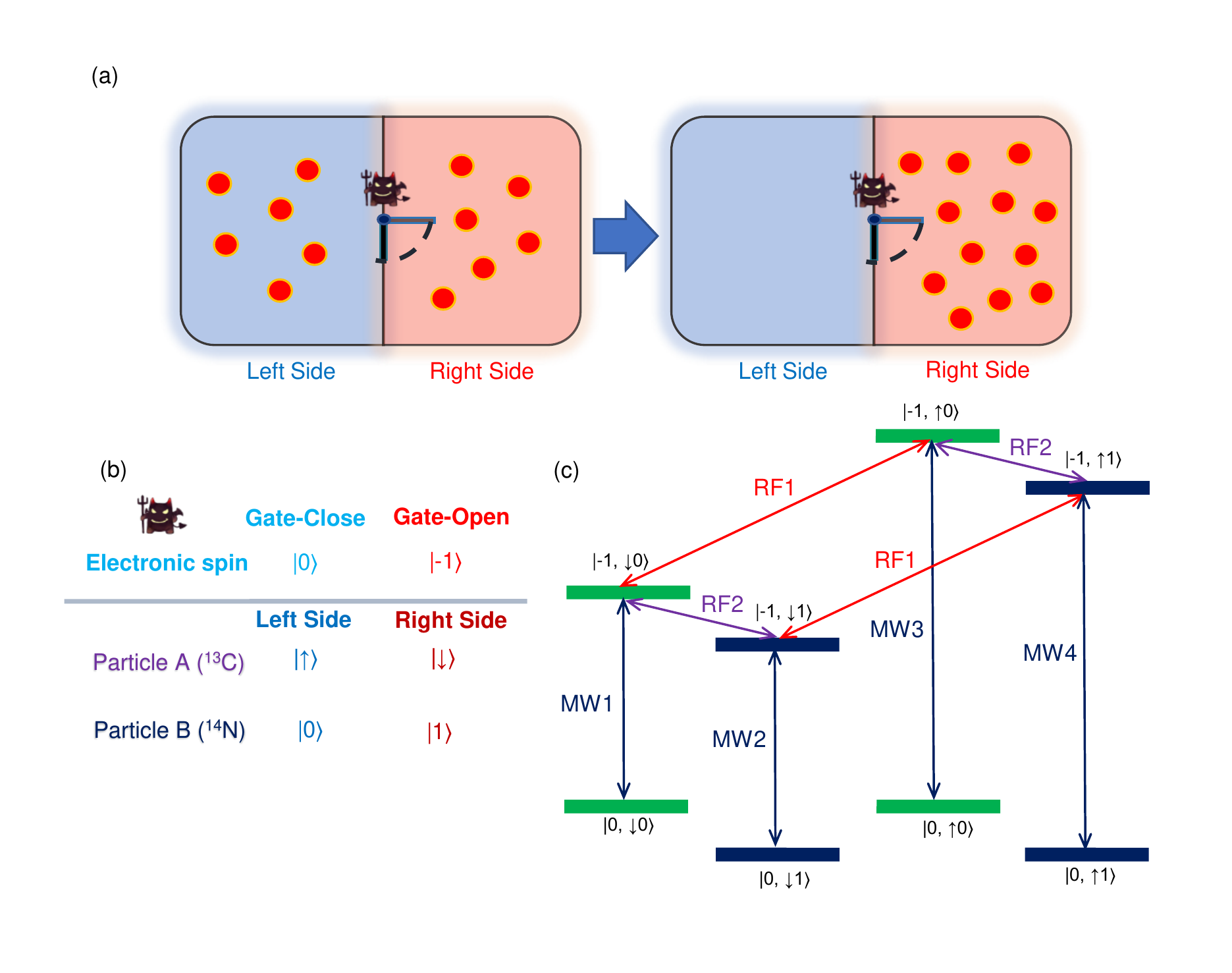}
\caption{\textbf{Implementation of quantum Maxwell's demon with solid state
spins.} \textbf{a}, Illustration of the Maxwell's demon paradox. A box with
gas particles is separated into two sides. A demon at the middle detects
which side the particle comes from (information acquiring) and only opens
the gate if the particle is from the left side (feedback). Eventually all
the particles move to the right side and the system entropy decreases which
seemingly contradicts with the second law of thermodynamics. \textbf{b},
Implementation of Maxwell's demon with three solid-state quantum spins. The
demon's memory is represented by the electronic spin state, which controls
opening or closing of the gate. Two particles A and B are represented by
nuclear spins ${^{13}}C$ and ${^{14}}N$, where the left or right sides of
the particles are encoded with two orthogonal spin states. The information
acquiring and feedback operations by the demon are implemented through
conditional quantum gates. \textbf{c}, The hyperfine level structure of the
electronic and the nuclear spins. The hyperfine interactions and the applied
external magnetic field spilt all the energy levels and differentiate their
transition frequencies. Based on the frequency selection, the applied
microwave or RF fields selectively drive a certain set of transitions and
make the conditional quantum gates. }
\end{figure}

To implement quantum Maxwell's demon, we use electronic spin and nuclear
spins in a nitrogen vacancy (NV) center of a room-temperature diamond sample
to carry the states of the system and the demon. The spins in diamond NV\
centers provide a promising solid-state system for realization of quantum
information processing and nanoscale sensing, which has attracted a lot of
interest in recent years \cite{24,25,26,27,28,29,30,31,32,33}. The ground
state of the NV center has electronic spin $S=1$, which interacts with the
dopant ${^{14}}N$ with nuclear spin $I=1$ and the nearby ${^{13}}C$ isotopes
with nuclear spin $I=1/2$ in the host diamond through hyperfine coupling,
forming a small quantum spin register \cite{24,25}. As shown in Fig. 1a and
1b, in our experiment we use the $|0\rangle $ and $|$-$1\rangle $ states of
the NV\ electronic spin to represent the demon's status for gate closing and
opening, respectively. In the first experiment, we use the states of $%
|\uparrow \rangle $ and $|\downarrow \rangle $ of a ${^{13}}C$ nuclear spin
with the hyperfine splitting of $6.5$ MHz to represent the location of the
first particle (particle A) in the left or right side of the box. To test
successive operation of the demon, we introduce the second particle
(particle B), whose location is represented by the $|0\rangle $ and $%
|1\rangle $ states of the ${^{14}}N$ nuclear spin with the energy splitting
about $2.8$ MHz. The relevant level structure of the electronic and the
nuclear spins in the NV center is Fig. 1c, and the transitions between these
levels can be manipulated through application of microwave or radio
frequency (RF) pulses at the corresponding frequencies.

The operation of Maxwell's demon has two components. The first component is
information acquiring which changes its memory state depending on the side
of incoming particle and in our case is realized through the quantum
controlled-NOT gate C$_{\text{C}}$-NOT$_{\text{E}}$ or C$_{\text{N}}$-NOT$_{%
\text{E}}$, where the electronic spin state representing the demon's memory
flips conditional on the control particle (${^{13}}C$ or ${^{14}}N$ nuclear
spin). The second component is a feedback operation where the demon opens
the gate conditional on its memory in the $|1\rangle $ state and the
incoming particle changes its side in the box. In our case, this feedback
operation is realized through the conditional quantum gate C$_{\text{E}}$-NOT%
$_{\text{C}}$ or C$_{\text{E}}$-NOT$_{\text{N}}$ for the first and the
second particles, respectively. The quantum circuit that implements the
Maxwell's demon operations is shown in Fig. 2a.

For Maxwell's demon experiment, the system starts at a mixed state where the
particles have equal probabilities to be in the left or right side of the
box while the demon's memory is at a pure state so that it can track the
particle's information. In our experiment, we apply a weak external magnetic
field of $38$ Gauss to split the energy levels $|1\rangle $ and $|$-$%
1\rangle $ of the electronic spin, and the field is small enough so that the
nuclear spins remain at the completely mixed states when we polarize the
electronic spin to the $|0\rangle $ state through optical pumping. The CNOT\
gates for the demons's information acquiring operation are implemented
through two microwave pulses and for the demons's feedback operation are
implemented through a RF pulse. For instance, with the level structure shown
in Fig. 1c, application of $\pi $-pulses with MW3 and MW4 (MW2 and MW4)
realize the C$_{\text{C}}$-NOT$_{\text{E}}$ (C$_{\text{N}}$-NOT$_{\text{E}}$%
) gate, and application of $\pi $-pulses with RF1 (RF2) realize the C$_{%
\text{E}}$-NOT$_{\text{C}}$ (C$_{\text{E}}$-NOT$_{\text{N}}$) gate. The
typical experimental sequence is shown in Fig. 2b.

The essence of Maxwell's demon is to reduce the entropy of the system at the
cost of entropy increase of its own memory. To characterize the entropy flow
for the system and the demon, we directly measure the entropy of the system
and the demon through quantum state tomography before and after the demon's
operation on the first and the second particles. Quantum state tomography is
implemented with a set of microwave and RF\ pulses that rotate the detection
basis and then optical read-out of the electronic spin state by detection of
fluorescence (see the Methods). For outcome states $i$ measured to occur
with probability $p_{i}$, the entropy is defined as $S=-\sum_{i}p_{i}\log
_{2}p_{i}$, where the base is taken to be $2$ for convenience. For
comparison between the experimental state (constructed as a density matrix $%
\rho _{\text{e}}$) and the outcome state $\rho _{\text{id}}$ in the ideal
case, we also measure the state fidelity $F$, which is defined as $F=\left[
tr\left( \sqrt{\sqrt{\rho _{\text{id}}}\rho _{\text{e}}\sqrt{\rho _{\text{id}%
}}}\right) \right] ^{2}$.

The measurement results for the state fidelity $F$ and the entropy $S$ are
shown in Fig. 2c and 2d. Before the demon's operation, both particles
(nuclear spins) are in the maximally mixed states with entropy $S=1$. The
demon (electronic spin) is in a pure state with a high fidelity (our
calibration of the fluorescence levels for detection has subtracted the
initialization imperfection by the green laser, see the Methods), but its
entropy is measured to be $0.10_{-0.09}^{+0.10}$ as the value of $S$ is
extremely sensitive to small imperfection near a pure state by its
definition. In the supplementary information, we provide an analysis of
noise and plot the entropy as a function of the system noise and the purity of quantum state.
From the plot, one can immediately see the sensitivity of entropy under
small drop of the purity of quantum state. After the demon's operation on
the first particle, the system (particles A and B) entropy decreases as
expected while the entropy for the demon's memory increases. Due to the
sensitivity of $S$ to small imperfection, the measured value of $S$ for
particle A is $0.36_{-0.05}^{+0.04}$ although it has a high fidelity of $%
95.5_{-0.8}^{+0.8}\%$ to a pure state, where the numbers in the superscript
and the subscript denote a confidence interval with probability of $68\%$.
We then continue to apply the demon's operation on the second particle. As
the demon has run out of its memory (which is only $1$ bit in this
experiment), the successive operation of the demon on the second particle
does not cause any entropy decrease for the system as one can see from the
measurement result in Fig. 2d. This confirms that the consumption of
information storage capacity in the demon's memory is essential for its
operation, a key concept for the resolution of Maxwell's demon paradox with
the second law of thermodynamics.

From the above data, one can see that the entropy decrease in the system is
always bounded by the information acquired by the demon during the readout
step, a key observation in interpretation of the Maxwell's demon experiment.
For instance, for the demon's interaction with particle A, the entropy
decrease in the system is $0.58$. while the information acquired by the
demon is given by $0.81$. For the demons's interaction with particle B,
although the operations are the same, as the demon runs out of its memory,
the information acquired by the demon is diminishing. So this limits the
entropy decrease of the particle B by the demon's operation, measured to be $%
0$ within the experimental uncertainty from Fig. 2.

In our experiment, the measurements are focused on detection of the entropy
flow for two reasons: first, the entropy plays a critical role for
interpretation of the Maxwell's demon paradox and its resolution with the
second law of thermodynamics. Second, in our system it is more convenient to
measure the entropy flow, which is complementary to the demonstration of
classical Maxwell's demon in other systems where it is typically easier to
measure the energy extraction instead of the entropy flow \cite%
{9,9b,10,11,12}. Although we cannot directly measure the work extraction in
this experimental system, the entropy decrease is closely related to the
amount of work that can be extracted from the system as discussed in Ref.
\cite{6}. In principle, the work could be extracted through stimulated
emission by shining a microwave pulse resonant to this spin although in
practice it is hard to realize that experimentally on a single spin. The
measured entropy decrease provides an upper bound on the amount of work that
could be extracted from the system through the Landauer's principle \cite%
{2,3,4,5,6,7}.

\begin{figure}[tbp]
\centering
\includegraphics [scale = 0.7]{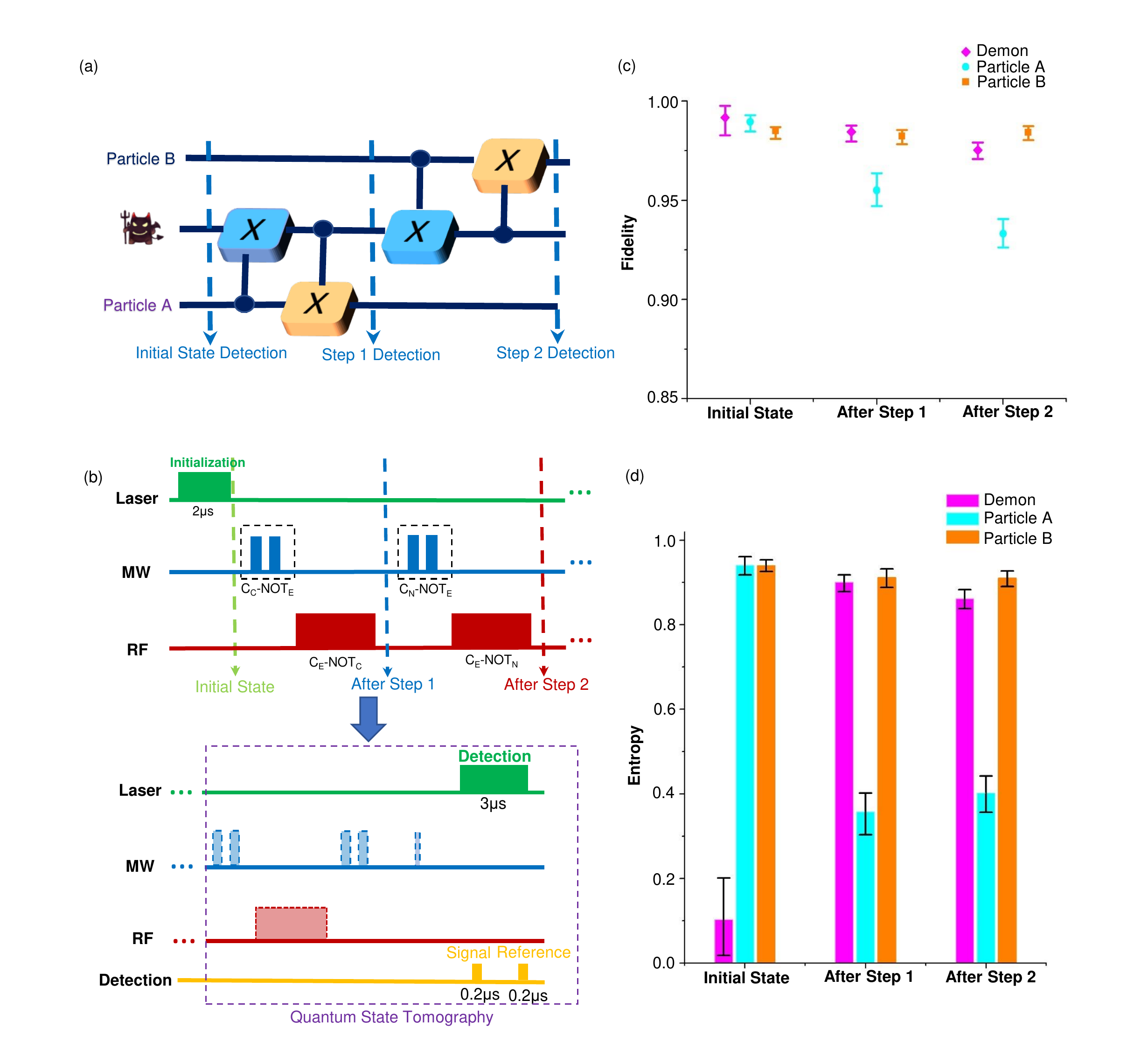}
\caption{\textbf{Maxwell's demon under successive operations.} \textbf{a},
The quantum circuit represents the demon's operations successively on the
first (A) and the second (B) particles. Each demon's operation consists of
two controlled-NOT quantum gates respectively for information acquiring and
feedback. We detect the state of the demon, particle A, and particle B at
the beginning and after the first (step 1) and the second (step 2) demon's
operation. \textbf{b}, Experimental sequence of laser, microwave, and RF
pulses for implementation of the successive demon's operations on particles
A and B. Each conditional gate on the electronic spin is implemented by two
microwave pulses on the corresponding transitions. The conditional gates on
the nuclear spins are realized through RF pulses. The detections at the
beginning, step 1, and step 2 are done through quantum state tomography,
which consists of several microwave and RF pulses to rotate the measurement
bases followed by the optical readout of the electronic spin state (see the
Methods). \textbf{c}, The measured state fidelity of the demon and particles
A and B at the beginning and after step 1 and step 2 compared with the ideal
states at the corresponding stages. \textbf{d}, The entropy of the system
and particles A and B measured through quantum state tomography at the three
corresponding stages. The error bars in all the figures denote the
confidence interval with probability of $68\%$, which corresponds to one
standard deviation if the underlying distribution is Gaussian.}
\end{figure}

\subsection{Maxwell's demon in a quantum superposition state}

\begin{figure}[tbp]
\centering
\includegraphics [scale = 0.7]{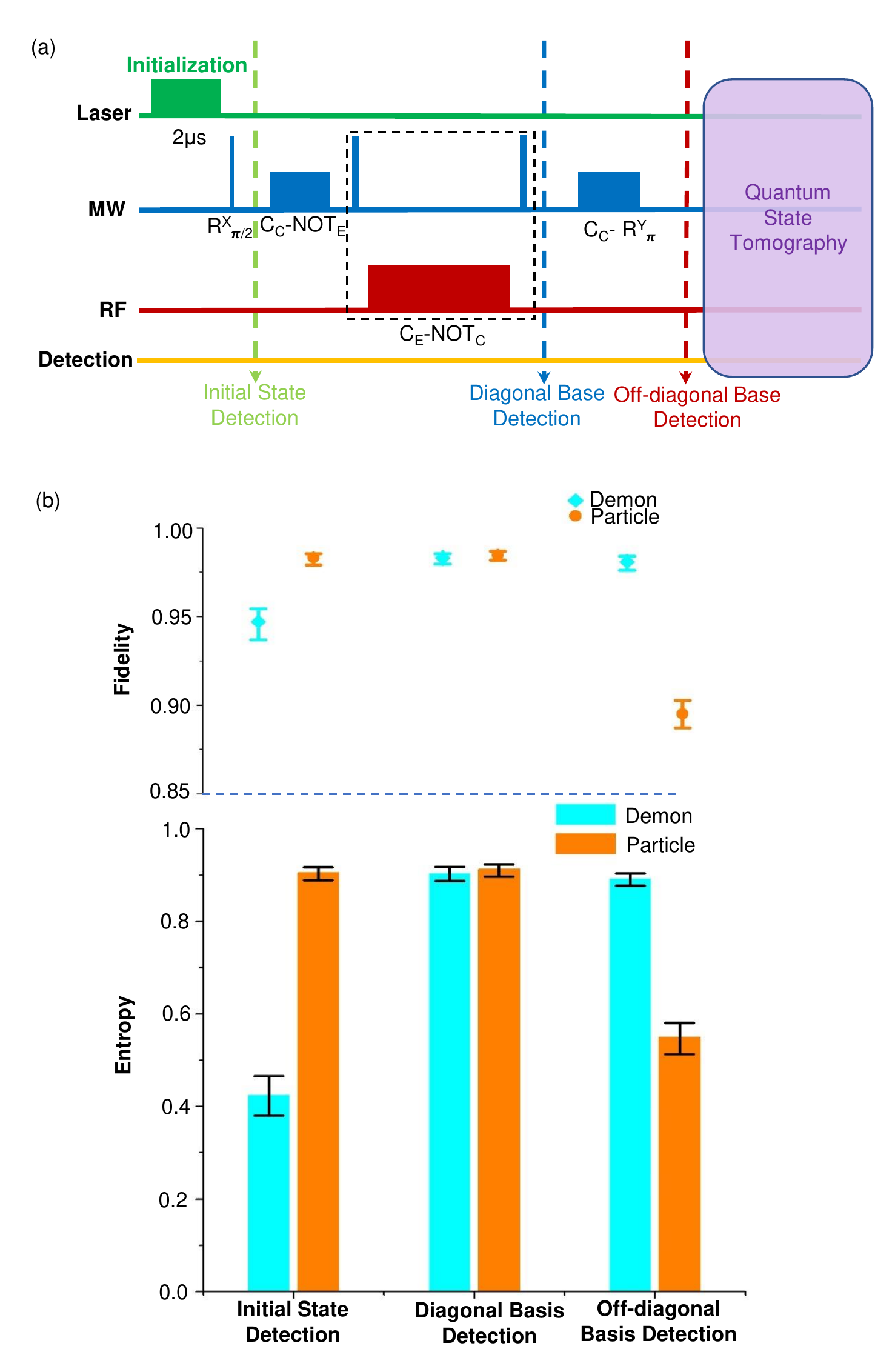}
\caption{\textbf{Maxwell's demon in a quantum superposition state.} \textbf{a%
}, The experimental sequence of laser, microwave, and RF pulses. The first
microwave pulse $R_{\protect\pi/2}^{X}$ induces a $\protect\pi/2$ rotation
along the $X$ spin-axis and prepares the demon in an equal superposition
state. The demon's operation between the electronic spin and the ${^{13}}C$
nuclear spin is achieved through the same conditional quantum gates, with
the only difference of addition of two dynamical decoupling $\protect\pi$%
-pulses to remove the dephaing noise of the electronic spin during the slow
RF pulse on the nuclear spin. We detect the output state in both the
diagonal (computational) and off-diagonal bases. For detection in the
off-diagonal basis, we fist apply a conditional gate (a microwave pulse $%
C_C-R_{\protect\pi}^{Y}$) to disentangle the electronic and the nuclear
spins as the demon's operation entangles them when the demon starts in a
superposition state. \textbf{b}, The measured state fidelity and entropy at
the beginning and after the demon's operation. The final state is detected
in both the diagonal and the off-diagonal bases. The fidelity denotes the
comparison between the experimental states and the ideal states at the
corresponding stages.}
\end{figure}

Next we prepare Maxwell's demon (the electronic spin in our case) initially
at a quantum superposition state $(|0\rangle +|1\rangle )/\sqrt{2}$ and test
influence of quantum coherence to the demon's operation. For this
experiment, we use the ${^{13}}C$ nuclear spin as our system particle, which
initially needs to be in a mixed state. However, it is better to polarize
the ${^{14}}N$ nuclear spin otherwise it works as a decoherence source for
the electronic spin when the latter is in a superposition state. We
therefore tune the external magnetic field to a value of $340$ Gauss, and
under this value of magnetic filed, the ${^{14}}N$ nuclear spin has a strong
enough spin-exchange interaction with the electronic spin at the excited
state so that it gets polarized by the optical pumping, while the ${^{13}}C$
nuclear spin only undergoes a weak spin-exchange interaction and remains to
be mixed. The demon's operation is achieved by the same set of quantum
gates. The only difference is that we need to apply dynamical decoupling
pulses, which induce a fast $\pi $ rotation along the X or Y spin axis, to
the electronic spin before and after the long RF pulse for the C$_{\text{E}}$%
-NOT$_{\text{C}}$ gate to remove the coupling of the electronic spin to the
spin bath and keep its coherence. The experimental pulse sequence is shown
in Fig. 3a. After the demon's operation, if we measure the system in the
classical basis (the computational basis $|\uparrow \rangle $ and $%
|\downarrow \rangle $), the system entropy does not show any decrease. This
can be easily understood as a demon in an equal quantum superposition of
classical states is just like a maximally mixed state looking from a
measurement in the classical basis. However, if we measure the system in the
superposition basis, the system entropy does decrease. The state fidelity
and the measured entropy under different bases are shown in Fig. 3b and 3c.
So when the demon hides its information storage capacity in the quantum
superposition state, maintaining coherence is important for functioning of
the quantum demon.

\subsection{Maxwell's demon in an entangled state with an ancilla}

\begin{figure}[tbp]
\centering
\includegraphics [scale = 0.7]{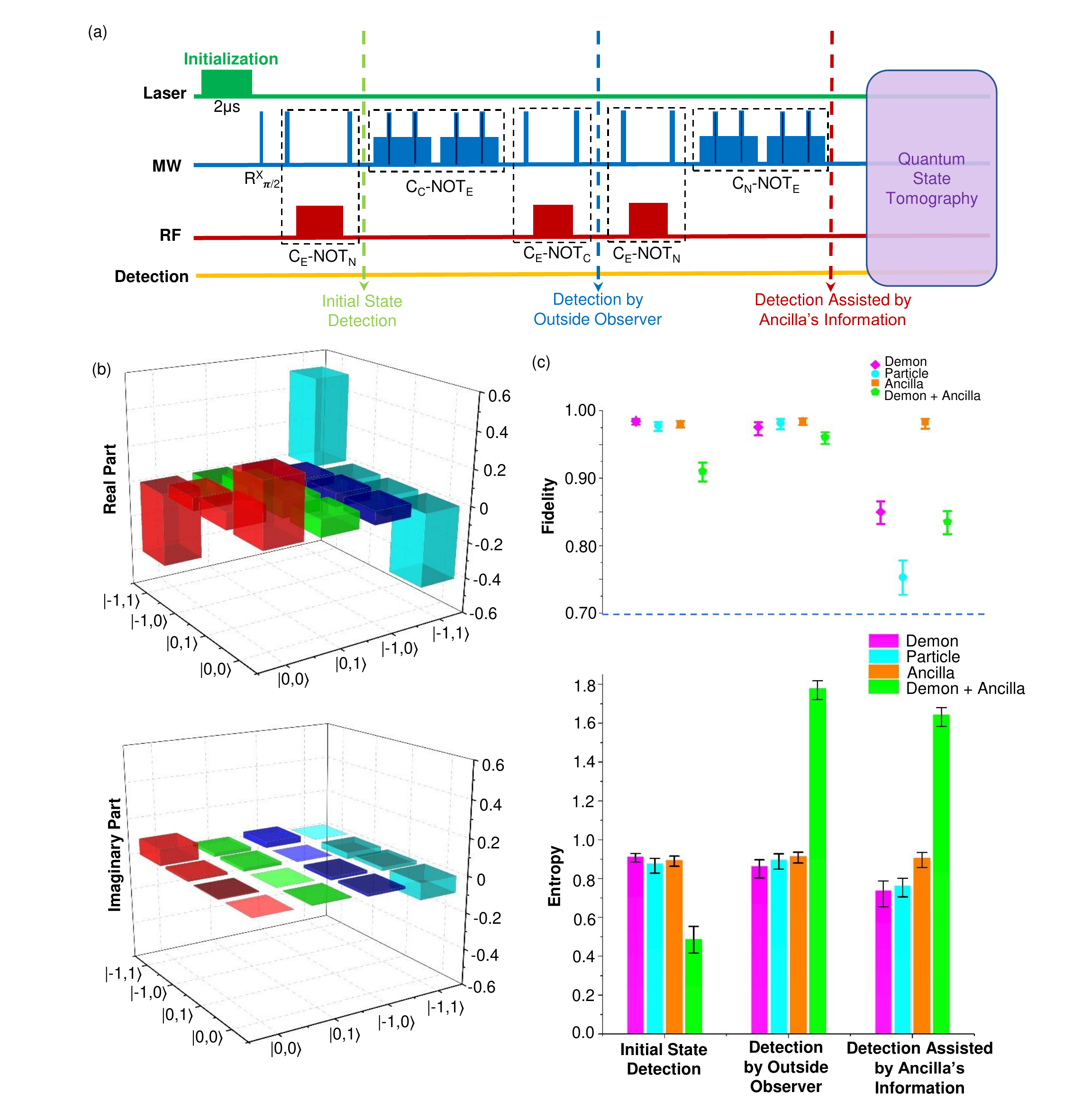}
\caption{\textbf{Maxwell's demon in an entangled state with an ancilla.}
\textbf{a}, The experimental Pulse sequence for initial entangled state
preparation, the demon's operation, and final state detection under
different scenarios. The first part prepares a maximally entangled state
$(|0,0\rangle-|\textnormal{-}1,1\rangle)/\protect\sqrt{2}$
between the demon (the electronic spin) and the ancilla (the ${^{14}}N$
nuclear spin). The demon's operation is the same as those in Fig. 3, with
the difference that now the microwave pulse becomes slow as one needs to
resolve the level splitting of the ${^{14}}N$ nuclear spin and the
electronic spin coherence during the slow microwave pulse is protected by
dynamical decoupling as well. For the final state detection, an outside
observer detects directly the demon (the electronic spin) and the system
(the ${^{13}}C$ nuclear spin), while an inside observer, with access to the
ancilla's information, applies two coherence-protected conditional quantum
gates between the electronic spin and the ${^{14}}N$ nuclear spin before
quantum state tomography on the system and the demon, and obtains results
inaccessible to the outside observer and incompatible with interpretation of
classical Maxwell's demons. The system and the demon are disentangled by a
microwave pulse (corresponding to a $C_{C}-NOT_E$ gate) in quantum state
tomography. \textbf{b}, Real and imaginary parts of the reconstructed
density matrix elements for the initial entangled state between the demon
and the ancilla. \textbf{c}, The measured state fidelity and entropy at the
beginning and after the demon's operation. The results for the final state
are shown for both the outside and the inside observers with and without
access to the ancilla's information, respectively. The notation "Demon +
Ancilla" corresponds to a detection of the joint state and the von Neumann
entropy for this composite system, while the others correspond to detection
of each individual system. }
\end{figure}

Entanglement represents the key feature of a quantum system, and it provides
new possibilities for a quantum Maxwell's demon. When we prepare the demon
in a maximally quantum entangled state with an ancilla, to an outside
observer who has no access to information in the ancilla, the demon appears
to be in a maximally mixed state and thus it has no information storage
capacity to reduce the system entropy by the demon's operation. However, an
inside observer who has access to the ancilla can make use of this
entanglement and find that the entropy in both the system and the demon can
decrease with feedback from the ancilla. This provides a nice demonstration
that entanglement can serve as a source of negative entropy for a quantum
thermal machine \cite{20}.

To demonstrate Maxwell's demon starting from an entangled state, we still
tune the external magnetic field to $340$ Gauss and polarize both the
electronic spin (the demon) and the ${^{14}}N$ nuclear spin (the ancilla) by
the optical pumping, which leaves the ${^{13}}C$ nuclear spin (the system)
in an almost maximally mixed state. Then, we can apply the microwave and RF\
pulses as shown in Fig. 4a to prepare the demon and the ancilla in a
maximally entangled state $|\Phi _{\text{DA}}\rangle =(|0,0\rangle -{|}${%
\textnormal{-}}${1,1\rangle })/\sqrt{2}$. The RF pulse for the gate C$_{%
\text{E}}$-NOT$_{\text{N}}$ is long, so the electronic spin coherence needs
to be protected by the dynamical decoupling pulses. Furthermore, this RF
pulse also induces a phase shift for the electronic spin by off-resonant
coupling, which is calibrated and compensated by the phase of subsequent
microwave pulses applied on the electronic spin. After the state
preparation, we have measured the entanglement between the demon and the
ancilla, with the result shown in Fig. 4b. From the reconstructed density
matrix $\rho _{\text{DA}}$ of the demon and the ancilla from quantum state
tomography, we find that the entanglement fidelity, defined as $F_{\text{e}%
}=\left\langle \Phi _{\text{DA}}\right\vert \rho _{\text{DA}}|\Phi _{\text{DA%
}}\rangle $, is $91.0_{-1.5}^{+1.3}\%$ for this initial state, which
confirms entanglement.

The demon's operation is achieved by the same set of gates as shown in Fig.
4a. The coherence of the electronic spin during the conditional gate C$_{%
\text{C}}$-NOT$_{\text{E}}$ needs to be protected by the dynamical
decoupling pulses, whose spin rotation axis is chosen to be the same as that
of the microwave pulses for the C$_{\text{C}}$-NOT$_{\text{E}}$ gate,
therefore the dynamical decoupling commutes with and does not affect the
gate operation. After the demon's operation, we immediately perform
measurements on the system and demon, emulating the measurement capability
of an outside observer who does not have access to the ancilla. In this
case, with the result shown in Fig. 4c, we find that the entropy of both the
system and the demon are at the maximum level and do not show any decrease
by the demon's operation. This is expected as without access to the ancilla,
the demon's entanglement information cannot be retrieved and it has no
memory capacity to reduce the system entropy.

We then emulate the capability of an inside observer and perform the
feedback operation between the demon and the ancilla shown in Fig. 4a. Same
as the demon operation, this is done by applying the gates C$_{\text{E}}$-NOT%
$_{\text{N}}$ and C$_{\text{N}}$-NOT$_{\text{E}}$ (with coherence protected
by the dynamical coupling). After the gates, we perform quantum state
tomography on the system and the demon (they are disentangled by a gate C$_{%
\text{C}}$-NOT$_{\text{E}}$), and the resultant entropy is shown in Fig. 4c.
For this inside observer, the entropies for the system and for the demon
both decrease. In the ideal case, the entropy should decrease to zero,
however, in experiments because of sensitivity of the value of entropy to
imperfections near a pure state, we find the entropy only decreases a bit
although the fidelity to a pure state is given by $85.0_{-1.8}^{+1.6}\%$ and
$75.3_{-2.6}^{+2.5}\%$ for the demon and the system, respectively. As the
ancilla's entropy remains almost the same before and after the demon's
operation, if we add up the total entropy of the system, the demon, and the
ancilla, they decrease by the demon's operation. This is impossible for a
classical demon, however, in the quantum case, with feedback from the
ancilla, we use one bit of entanglement, which can convert to two bits of
negative entropy \cite{20}, and therefore it is possible to reduce the
entropy for both the system and the demon. When there is entanglement, we
need to count entropy by a collective measurement on the demon and the
ancilla in an entangled basis instead of adding up their individual
entropies, and as shown in Fig. 4, the total entropy counted in this way
then increases and there is no inconsistency with the second law of
thermodynamics by taking into account of the contribution of entanglement.

\section{Discussion}

We have implemented the Maxwell's demon experiment using a quantum system
with three solid-state spins in a diamond sample. Our experiment
demonstrates for the first time the role of quantum entanglement in the
Maxwell's demon. We have prepared the demon in a quantum entangled state
with an ancilla and demonstrated that the entanglement provides a resource
of negative entropy for the system and the demon if the observer has access
to the ancilla's information \cite{20}. Through successive application of
the demon's operation on system particles, we have also demonstrated that
the memory capacity of Maxwell's demon is critical for its functioning, an
observation important for the resolution of this paradox with the second law
of thermodynamics \cite{2,3,4,5,6}.

A quantum implementation of Maxwell's demon, in particular with access to
the entangled states, deepens our understanding of the interplay between
thermodynamics and quantum entanglement and sheds new light on the profound
connection between the fundamental principles of thermodynamics and quantum
information theory \cite{6,7,8,18,19,20,21}. The quantum control techniques
in the Maxwell's demon experiment may also find applications in
implementation of other quantum thermal machines involving microscopic
systems \cite{18,19,20,21}.

\section{Methods}

\subsection{Experimental setup}

We use a home-built confocal microscopy with an oil immersed objective lens
to address and detect single nitrogen vacancy (NV) centers in a
single-crystal diamond sample, which is mounted on a doughnut shape,
three-axis, closed-loop piezoelectric actuator with sub-micron resolution. A
$532$ nm diode laser, controlled by an acoustic optical modulator (AOM), is
used for spin state initialization and detection. Fluorescence photons
(wavelength ranging from $637$ to $850$ nm) are collected into a single-mode
fibre and detected by a single photon counting module (SPCM). To deliver the
microwave signal to the NV center, we use a gold coplanar waveguide (CPW),
which has a small $\Omega $-shape structure (of $60$ $\mu $m length) near
the NV\ to enhance the coupling. To deliver the RF signals to the NV center,
we set a gold coil above the sample. The coil has $0.5$ cm diameter and 50
turns, with $20$ $\mu m$ gaps between the gold lines.

We scan the sample and find a single NV center coupled with a proximal ${%
^{13}}C$ nuclear spin of $6.5$ MHz hyperfine coupling strength for our
experiment. The external magnetic field is applied through a permanent
magnet, which is tuned to $38$ G along the NV axis for the first experiment
and to $340$ G for the second and third experiments. We use an
arbitrary-waveform generator (AWG, Tektronix with $1$ GHz sampling rate) to
control the time sequence of our experiment. The digital markers of the AWG
are used to control the pulse sequence with a timing resolution of $1$ ns.
The RF\ signals for control of nuclear spins and the digital signals for
switch of the photon counters and the home-built field programmable gate
array (FPGA) circuit are directly generated from the AWG. The microwave
signals from a generator (MXG Analog Signal Generater, Keysight with
9kHz-6GHz frequency) are mixed with the control signals from the AWG via IQ
mixers. All the microwave and RF signals are amplified by independent
amplifiers.

For each experimental cycle, We collect signal photons for $200$ ns right
after the detection laser rises and reaches the full intensity, and for
another $200$ ns for reference $2$ $\mu $s later. We repeat the experimental
cycle at least $4\times 10^{6}$ times, resulting in a total photon count
over $5\times 10^{4}$. The error bars of our data describe the statistical
error which comes from the photon counting. To calculate the error bar for
each experimental quantity, we use Monte Carlo sampling by assuming a
Poissonian distribution for the photon counts and propagate the statistical
distribution from the measured data to the calculated quantity through exact
numerical simulation. This simulation based on Monte Carlo sampling gives
the statistical distribution for the target quantity, from which we can
easily calculate the confidence intervals. The error bars in the paper
denote an confidence interval with probability of $68.3\%$, which
corresponds to one standard deviation if the underlying distribution is
Gaussian.

\subsection{Experimental calibration and quantum state tomography}

The fluorescence levels of electronic spin $\left\vert 0\right\rangle $ and $%
\left\vert \text{\textnormal{-}}1\right\rangle $\ states, as well as the
contrast between them, vary slightly for different experimental cycles due
to temperature variation or change of the laser power. To cancel out the
influence of this variation, before each measurement sequence we calibrate
the fluorescence levels of the electronic spins in $\left\vert
0\right\rangle $ and $\left\vert \text{\textnormal{-}}1\right\rangle $
states and use this as the standard to normalize our experimental data in
the measurement sequence. For this calibration, we have neglected the
imperfection of microwave $\pi $-pulse to flip between the electronic spin $%
\left\vert 0\right\rangle $ and $\left\vert \text{\textnormal{-}}%
1\right\rangle $ states. By this calibration, we have subtracted the state
initialization and detection errors in the computational basis. The state
infidelities reported in the paper comes from the accumulated errors of a
sequence of quantum gates used for the demon operations and quantum
entanglement as well as the decoherence contribution during this sequence of
quantum gates.

We use quantum state tomography to measure the density matrix of the final
state. The electronic spin state is directly measured by its corresponding
fluorescence levels together with microwave pulses to rotate the measurement
basis. The nuclear spin states for ${^{13}}C$ or ${^{14}}N$ are first mapped
to the electronic spin states with a microwave pulse at the appropriate
frequency for measurement of their fluorescence levels, with the calibration
procedure same as those described in the method section of Ref. \cite{32}.
For the ${^{14}}N$ nuclear spin, we only use its $\left\vert 0\right\rangle $
and $\left\vert 1\right\rangle $ states to represent the qubit, and its
population in the $\left\vert \text{\textnormal{-}}1\right\rangle $ level,
when exists, only contributes to the background contrast in calibration of
the fluorescence. The rotation of the nuclear spin measurement basis is
achieved through a RF\ pulse at the corresponding frequency.

\subsection{State transformation in the third experiment}

Here, we work out the step-by-step state transformation for the third
experiment where the demon (the electronic spin) starts in an entangled
state of the form $|\Psi _{0}\rangle _{\text{EN}}=(|0\rangle _{\text{E}%
}|0\rangle _{\text{N}}-|$\textnormal{-}$1\rangle _{\text{E}}|1\rangle _{%
\text{N}})/\sqrt{2}$ with the ancilla (the nitrogen nuclear spin). Note that
the reduced initial state of the demon is given by a fully mixed $2\times 2$
identity matrix $I_{\text{E}}$. The system (the C$^{13}$ nuclear spin) is
initially prepared in a full mixed state described by the identity matrix $%
I_{\text{C}}$ (this C$^{13}$ nuclear spin will not be polarized by optical
pumping under the $340$ G magnetic field as verified in the experiment).
After the demon's operation C$_{\text{E}}$-NOT$_{\text{C}}$C$_{\text{C}}$-NOT%
$_{\text{E}}$, the state transforms to $\rho _{\text{ECN}}=(|\Phi
_{1}\rangle \left\langle \Phi _{1}\right\vert +|\Phi _{2}\rangle
\left\langle \Phi _{2}\right\vert )/2$, where $|\Phi _{1}\rangle =\left(
|0\rangle _{\text{E}}|\uparrow \rangle _{\text{C}}|0\rangle _{\text{N}}+i|%
\text{-}1\rangle _{\text{E}}|\downarrow \rangle _{\text{C}}|1\rangle _{\text{%
N}}\right) /\sqrt{2}$ and $|\Phi _{2}\rangle =\left( |0\rangle _{\text{E}%
}|\downarrow \rangle _{\text{C}}|1\rangle _{\text{N}}+i|\text{-}1\rangle _{%
\text{E}}|\uparrow \rangle _{\text{C}}|0\rangle _{\text{N}}\right) /\sqrt{2}$%
. For an outside observer who does not have access to the information of the
ancilla, the system and demon's state is described by the reduced density
matrix $I_{\text{C}}\otimes I_{\text{E}}$. This is a completely mixed state
and the entropy in either the system or the demon does not show any
decrease, as confirmed by the experimental observation. However, for an
inside observer who has access to the ancilla, the observer can apply joint
operations C$_{\text{N}}$-NOT$_{\text{E}}$C$_{\text{E}}$-NOT$_{\text{N}}$ on
the ancilla and the demon, which transforms the state to the form $\left[
(|0\rangle _{\text{E}}|\uparrow \rangle _{\text{C}}{+}|1\rangle _{\text{E}%
}|\downarrow \rangle _{\text{C}})/\sqrt{2}\right] \otimes I_{\text{N}}$. The
reduced state for the system and the demon becomes $(|0\rangle _{\text{E}%
}|\uparrow \rangle _{\text{C}}+|1\rangle _{\text{E}}|\downarrow \rangle _{%
\text{C}})/\sqrt{2}$. After another disentangling operation C$_{\text{C}}$%
-NOT$_{\text{E}}$, both the system and the demon are in the pure state $|$%
\textnormal{-}$1\rangle _{\text{E}}\otimes (|\uparrow \rangle +|\downarrow
\rangle )_{\text{C}}/\sqrt{2}$, and the entropy for either of them decreases
by one bit in the ideal case. If we compare this case with the measurement
by an outside observer (or with the initial state), we see that the
entanglement with the ancilla effectively contributes two bits of negative
entropies to the system and the demon with the ancilla assisted operations,
reducing the entropy of each of them by one bit. This is similar in spirit
to the scheme in Ref. \cite{20}. For the experimental case, due to the
sensitivity of entropy to imperfection near pure states and the sensitivity
of entanglement to noise during a series of quantum operations, we observed
an entropy decrease about $0.12$ bit and $0.14$ bit for the system and the
demon, respectively. In the supplementary information, we take into account
the contribution of dominant noise during the experiment, and the
theoretical estimation agrees semi-quantitatively with the experimental
observation. The observed entropy decrease, although significantly less than
the amount predicted for the ideal case, still confirms the contribution of
entanglement as a resource of negative entropy, as otherwise the total
entropy for the system and the demon can only be increasing due to the
second law of thermodynamics, in particular with the noise contribution
during the experiment.

\textbf{Acknowledgements} This work was supported by the Ministry of Education and the National key Research
and Development Program of China 2016YFA0301902. LMD and ZYZ acknowledges in addition support from the MURI and the ARL CDQI program.

\textbf{Author Contributions} L.M.D. designed the experiment and supervised
the project. W.B.W, X.Y.C., F.W., P.Y.H., Y.Y.H., W.G.Z., X.L.O., X.Z.H.,
Z.Y.Z., L.H. performed the experiment. L.M.D. and W.B.W. wrote the manuscript.

\textbf{Author Information} The authors declare no competing financial
interests. Correspondence and requests for materials should be addressed to
L.M.D. (lmduan@umich.edu).


\section*{References}

\begin{thebibliography}{99}
\bibitem{1} Maruyama, K., Nori, F. \& Vedral, V. Colloquium: The physics of
Maxwell's demon and information. Rev. Mod. Phys. 81, 1--23 (2009).

\bibitem{2} Landauer, R. \& IBM, J. Res. Dev. 5, 183 (1961).

\bibitem{3} Bennett, C. The thermodynamics of computation---a review. Int.
J. Theor. Phys. 21, 905--940 (1982).

\bibitem{4} Zurek, W. H. in Frontiers of Nonequilibrium Statistical Physics
Volume 135 of Nato Science Series B: (eds. Moore, G. T. \& Scully, M. O.)
151 (Plenum Press, 1984).

\bibitem{5} Zurek, W. H. Thermodynamic cost of computation, algorithmic
complexity and the information metric. Nature 341, 119-124 (1989).

\bibitem{6} Lloyd, S. Quantum-mechanical Maxwell's demon. Phys. Rev. A 56,
3374--3382 (1997).

\bibitem{7} Vedral, V. Landauer's erasure, error correction and
entanglement. Proc. R. Soc. London, Ser. A 456, 969. (2000).

\bibitem{8} Kim, S., Sagawa, T., De Liberato, S. \& Ueda, M. Quantum Szilard
Engine. Phys. Rev. Lett. 106, 70401 (2011).

\bibitem{9} Scully, M. O., Zubairy, M. S., Agarwal, G. S. \& Walther, H.
Extracting Work from a Single Heat Bath via Vanishing Quantum Coherence.
Science 299, 862-864 (2003)

\bibitem{9b} Serreli, V., Lee, C. F., Kay, E. R., and Leigh, D. A. Molecular
Information Ratchet. Nature, 445. 523 - 527 (2007).

\bibitem{10} Raizen, M. G. Comprehensive Control of Atomic Motion. Science
324 (5933):1403--1406 (2009).

\bibitem{11} Toyabe, S., Sagawa, T., Ueda, M., Muneyuki, E. \& Sano, M.
Experimental demonstration of information-to-energy conversion and
validation of the generalized Jarzynski equality. Nat. Phys. 6, 988 (2010).

\bibitem{12} Berut, A. et al. Experimental verification of Landauer's
principle linking information and thermodynamics. Nature 483, 187--9 (2012).

\bibitem{13} Koski, J. V., Maisi, V. F., Sagawa, T. \& Pekola, J. P.
Experimental observation of the role of mutual information in the
nonequilibrium dynamics of a Maxwell demon. Phys. Rev. Lett. 113, 30601
(2014).

\bibitem{14} Koski, J. V., Maisi, V. F., Pekola, J. P. \& Averin, D. V.
Experimental realization of a Szilard engine with a single electron. PNAS
111, 13786 (2014).

\bibitem{15} Koski, J. V., Kutvonen, A., Khaymovich, I.M., Ala-Nissila, T.
\& Pekola, J.P. On-Chip Maxwell's Demon as an Information-Powered
Refrigerator. Phys. Rev. Lett. 115, 260602 (2015).

\bibitem{16} Vidrighin, M. D. et al. Photonic Maxwell's Demon. Phys. Rev.
Lett. 116, 1-7 (2016).

\bibitem{17} Elouard, C., Herrera-Marti, D., Huard, B. \& Auffeves, A.
Extracting Work from Quantum Measurement in Maxwell's Demon Engines. Phys.
Rev. Lett. 118, 260603 (2017).

\bibitem{18} Kieu. T. D. The second law, Maxwell's demon, and work derivable
from quantum heat engines. Phys. Rev. Lett. 93, 140403 (2004).

\bibitem{19} Quan, H., Wang, Y., Liu, Y., Sun, C. \& Nori, F. Maxwell's
Demon Assisted Thermodynamic Cycle in Superconducting Quantum Circuits.
Phys. Rev. Lett. 97, 180402 (2006).

\bibitem{20} Rio, L.D., Renner, R., Aaberg, J., Dahlsten, O. \& Vedral, V.
The thermodynamic meaning of negative entropy. Nature 474, 61 (2011).

\bibitem{21} Masanes, L. \& Oppenheim, J. A general derivation and
quantification of the third law of thermodynamics. Nature Communications 8,
14538 (2017).

\bibitem{22} Camati, P. A. et al. Experimental rectification of entropy
production by a Maxwell's Demon in a quantum system. Phys. Rev. Lett. 117,
240502 (2016).

\bibitem{23} Cotteta, N., Jezouina, S., Bretheaua, L., Campagne-Ibarcqa, P.,
Ficheuxa, Q., Andersb, J., Auffeves, A., Azouitd, R., Rouchond, P. \&
Huarda,  B. Observing a quantum Maxwell demon at work. Proc. Natl. Acad.
Sci. 114, 7561-7564 (2017).

\bibitem{24} Doherty, M. W. et al. The nitrogen-vacancy colour centre in
diamond. Physics Reports 528, 1-45 (2013).

\bibitem{25} Childress, L., Walsworth, R. \& M. Lukin. Atom-like crystal
defects: From quantum computers to biological sensors. Physics Today 67, 38.
(2014).

\bibitem{26} Jelezko, F. Gaebel, T., Popa, I. Gruber, A. \& Wrachtrup, J.
Observation of coherent oscillations in a single electron spin. Phys. Rev.
Lett. 92. 076401 (2004).

\bibitem{27} Neumann, P., Mizuochi, N., Rempp, F., Hemmer, P., Watanabe, H.,
Yamasaki, S., Jacques, V., Gaebel, T., Jelezko, F. \& Wrachtrup, J.
Multipartite entanglement among single spins in diamond. Science 320, 1326
(2008).

\bibitem{28} Jacques, V., Neumann, P., Beck, J., Markham, M., Twitchen, D.,
Meijer, J., Kaiser, F., Balasubramanian, G., Jelezko, F. \& Wrachtrup, J.
Dynamic polarization of single nuclear spins by optical pumping of
nitrogen-vacancy color centers in diamond at room temperature. Phys. Rev.
Lett. 102. 057403 (2009).

\bibitem{29} Yao, N.Y., Jiang, L., Gorshkov, A.V., Maurer, P.C., Giedke, G.,
Cirac, J.I. \& Lukin, M.D. Scalable architecture for a room temperature
solid-state quantum information processor. Nature Commun. 3, 800 (2012).

\bibitem{30} Van Der Sar, T., Wang, Z. H., Blok, M. S., Bernien, H.,
Taminiau, T. H., Toyli, D. M., Lidar, D. A., Awschalom, D. D., Hanson R. \&
Dobrovitski, V. V. Decoherence-protected quantum gates for a hybrid
solid-state spin register. Nature 484, 82 (2012).

\bibitem{31} Maurer, P. C., Kucsko, G., Latta, C., Jiang, L., Yao, N. Y.,
Bennett, S. D., Pastawski, F., Hunger, D., Chisholm, N., Markham, M.,
Twitchen, D. J., Cirac, J. I., Lukin, M. D.. Room-temperature quantum bit
memory exceeding one second. Science 336, 1283 (2012).

\bibitem{32} Zu, C., Wang, W.-B., He, L., Zhang, W.-G., Dai, C.-Y., Wang, F.
\&  Duan, L.-M. Experimental realization of universal geometric quantum
gates with solid-state spins. Nature 514, 72 (2014).

\bibitem{33} Pfaff, W. et al. Unconditional quantum teleportation between
distant solid-state quantum bits. Science 10.1126, 1253512 (2014).
\end{thebibliography}
\end{document}


\title{Supplementary information: Realization of quantum Maxwell's demon \\
with solid-state spins}
\author{W.-B. Wang$^{1}$, X.-Y. Chang$^{1}$, F. Wang$^{1}$, P.-Y. Hou$^{1}$,
Y.-Y. Huang$^{1}$, W.-G. Zhang$^{1}$, X.-L. Ouyang$^{1}$, X.-Z. Huang$^{1}$,
Z.-Y. Zhang$^{2}$, L. He$^{1}$, L.-M. Duan}
\affiliation{Center for Quantum Information, IIIS, Tsinghua University, Beijing 100084,
PR China}
\affiliation{Department of Physics, University of Michigan, Ann Arbor, Michigan 48109, USA}
\maketitle


\section{Supplementary note 1: Analysis of experimental noise and its
influence}

First, we show that the entropy is significantly more sensitive to the
purity of a quantum state under small noise compared with the state
fidelity. Our experiment concerns about the entropy of a qubit system, and
the purity of quantum state of a qubit can be quantified by the length of
the Bloch vector for the state. The qubit sate can always be represented by
a Bloch sphere and the Bloch vector length $L$, defined as $L=\left\vert
2p-1\right\vert $, where $p$ and $1-p$ are the two eigenvalues of the
density matrix of a single-qubit state. Apparently pure states have $p=1$
and $L=1$. The relation between the Bloch vector length $L$ and the von
Neumann entropy $S$ (or the Shannon entropy in the diagonal basis) is given
by
\begin{eqnarray}
S &=&(-p\log _{2}p-(1-p)\log _{2}(1-p))  \notag \\
&=&1-\frac{1}{2}[(1+L)\log _{2}(1+L)+(1-L)\log _{2}(1-L)].
\end{eqnarray}%
This relation is shown in the supplementary figure 1a. First, the curve is
monotonically decreasing, so the entropy always increases when the purity
decreases as one expects. Second, the entropy sharply rises when the purity
reduces from the maximum value $1$ (as one deviates from the pure state) by
the nonlinear dependence. For instance, a $10\%$ drop in $L$ corresponds to
an increase of the entropy to $30\%$ of its maximum value already. This is
different from the state fidelity $F$, which is linearly dependent on the
deviation $1-L$ ($F\simeq p=1-\left( 1-L\right) /2$). This explains why the
entropy is significantly more sensitive to noise compared with the state
fidelity.

Our third experiment has the largest contribution from noise as we start
from an entangled state which is more sensitive to noise including dephasing
and the experimental sequence of operations is also significantly longer
compared with the first two experiments. We identify two major noise
contributions to the entropy increase in our system. First, the electronic
spin is subject to remaining dephasing although we have implemented
decoherence protected quantum gates using the dynamical decoupling pulses.
At room temperature, even after correction by dynamical decoupling pulses,
the coherence time for the electronic spin is still limited to be well below
$1$ ms. For instance, in the supplementary figure 1b, we show the detected
coherence decay of the electronic spin after application of the Hahn spin
echo in the between, with the effect measured by both the decay of the
purity $L$ and the increase of the single-spin entropy $S$. The decay of the
Bloch vector length can be fit by the Gaussian decay envelop $exp(-(\tau
/T_{2})^{2})$, and this fitting gives an estimated $T_{2}\sim 378$ $\mu s$.
The operation time of the whole experimental sequence in our third
experiment, is about $190$ $\mu s$. If we take the above estimated $T_{2}$,
the Bloch length vector will drop by about $23.4\%$ after the experimental
sequence. The second major noise contribution to our third experiment lies
in the electronic spin dephasing during the time gap between two adjacent
decoherence protected quantum gates. In our experiment, each CNOT\ gate is
implemented with combination of two decoherence-protected Toffoli gates, and
each Toffoli gate is coherence-protected by two symmetric dynamical
decoupling pulses (see Fig. 4a of the main text). The experimental microwave
and RF pulses for the Toffoli gates always have finite rising and falling
times, so there are small time gaps between two adjacent Toffoli gates,
which are outside the coherence protection by the dynamical decoupling
pulses. The decoherence rate for the unprotected time window should be
estimated by $1/T_{2}^{\ast }$, where $T_{2}^{\ast }$ denotes the
free-induction decay time for the electronic spin without the spin echo. The
free induction decay time $T_{2}^{\ast }$ depends on the external magnetic
filed and could range from $1.7$ $\mu s$ to $3.6$ $\mu s$ as measured in
\cite{1,2}. For our experiment, we estimate $T_{2}^{\ast }\sim 2.5$ $\mu s$.
The total time gaps $t_{\Delta }$ between the adjacent Toffoli gates in our
whole experimental sequence adds up to about $1.2$ $\mu s$, and the
dephasing during those unprotected gaps will contribute another $20.6\%$
decay of the Bloch vector length through estimation by a similar Gaussian
decay envelop $exp(-(t_{\Delta }/T_{2}^{\ast })^{2})$. As a result, the
estimated remaining length of the Bloch vector becomes $(1-0.234)\times
(1-0.206)=0.608$, which would correspond to an entropy of $0.72$ (in
contract to the zero entropy in the ideal case). This value is close to what
we observed in the third experiment as shown in Fig. 4c for the main text.
For the initial state, the system should be in a maximally mixed state and
have entropy $S=1$. However, due to small imperfection in the experimental
calibration and preparation, we have a measured entropy about $0.90$ in our
experiment (see Fig. 4c, note that this corresponds to a pretty high
preparation fidelity about $98\%$). Combining the entropies estimated from
the above analysis, we have the entropy decrease $\Delta S=0.90-0.72=0.18$
for our third experiment, which roughly agrees with what we observed
experimentally.
\begin{figure}[tbp]
\centering
\includegraphics [scale = 0.7]{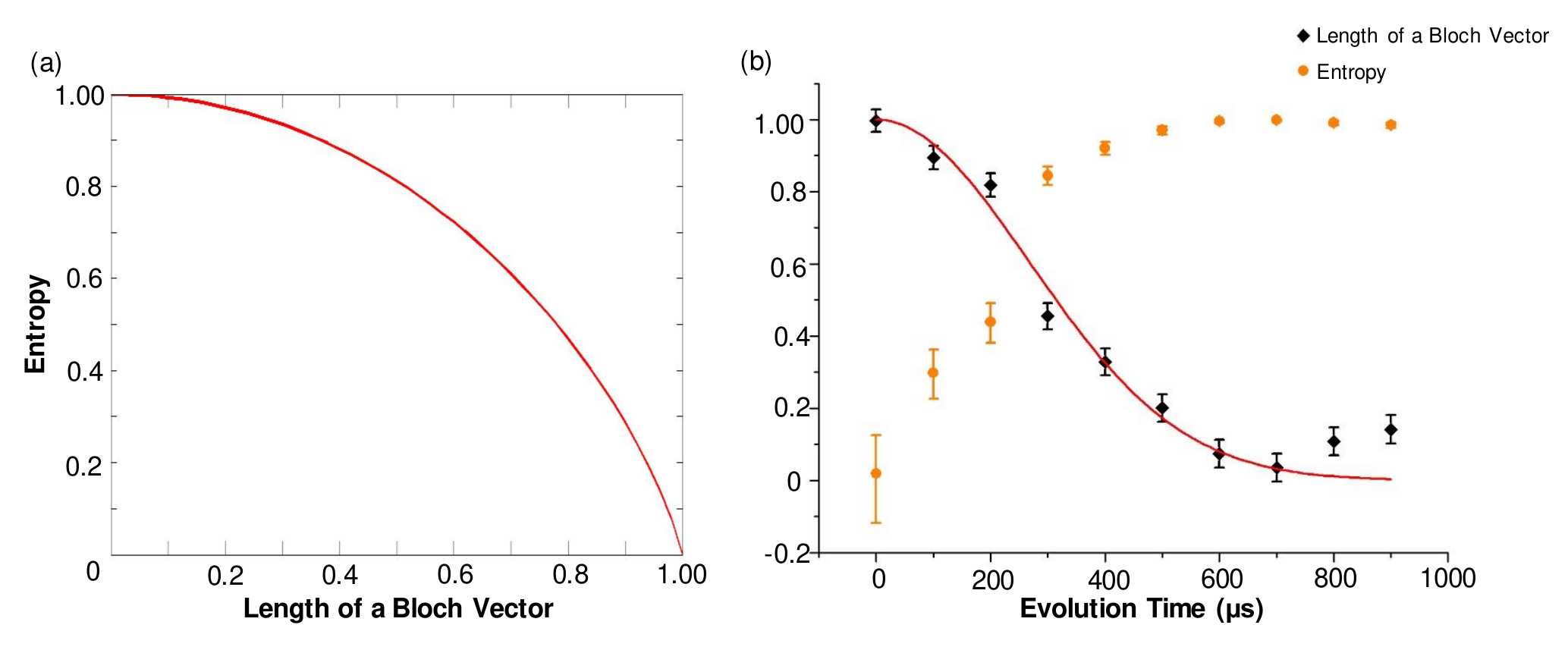}
\caption{\textbf{Entropy variation under experimental noise.} \textbf{a},
Variation of the entropy with the length of Bloch vector which characterizes the purity of 
a quantum state. It is given by a monotonic but nonlinear relation with the function $%
y=1-[(1+x)log_{2}(1+x)+(1-x)log_{2}(1-x)]/2$. One can see that entropy increases quickly when
the Bloch vector length deviates from $1$, explaining the sensitivity of entropy under small noise compared with
the state fidelity.  \textbf{b}, Decay of the length of Bloch vector under free evolution 
with a Haho echo (dynamical decoupling) pulse applied at the middle point of the evolution. 
The external magnetic field is set at $340$ Gauss. The data can be fit with a
Gaussian envelop $exp[(-\protect\tau /{T_{2}})^{2}]$ with
$T_{2}\sim 378$ $\protect\mu$s. }
\end{figure}

\section{Supplementary note 2: Influence of imperfect initial state
polarization}

In our experiment, we have used the same method as in Refs. \cite{3,3a} to
calibrate the fluorescence counts for different spin states and to normalize
our experimental results. This is a widely used calibration method in
diamond NV center experiments which effectively subtracts the influence of
the imperfection in the initial state polarization. To see how it works, we
describe the electronic spin state with imperfect initial polarization
by the density operator
\begin{equation}
\rho =p_{0}\rho _{0}+\left( 1-p_{0}\right) I_{2}/2
\end{equation}%
where $\rho _{0}$ denotes the effective initial state (usually pure), $I_{2}$
is the $2\times 2$ identity matrix, and $p_{0}$ denotes the polarization
probability, which is typically about $84\%$ for a diamond NV center at room
temperature \cite{4}. Any operations on the state $\rho $ leaves the second part $I_{2}/2$ unchanged,
and this is even true for general non-unitary operations. For the
fluorescence count measurements on the diamond NV center, we calibrate the
count contrast right after the state initialization and defines the
underlying state as $\left\vert 0\right\rangle $ ($\left\vert 1\right\rangle
$) when the count reaches the maximum (minimum) \cite{3,3a}. For the final
state measurement, we use this calibrated count contrast to measure the
state fraction in $\left\vert 0\right\rangle $ or $\left\vert 1\right\rangle
$ components. As the term $\left( 1-p_{0}\right) I_{2}/2$ in $\rho $ has no
influence on the count contrast, it only contributes a background noise and
we effectively subtract the influence of this term from both the initial and
the final states.

As we have subtracted the same identity term in the initial and the final
state by this normalization method, the observation of an entropy decrease
(or increase) should be independent of this background noise subtraction. {%
Note that the second term $\left( 1-p_{0}\right) I_{2}/2$ has zero contribution to the length of 
the Bloch vector, and the normalization factor $p_{0}$ in the first term $p_{0}\rho _{0}$ reduces
the Bloch vector length $L$ for the initial and the final states by the same ratio,
so it will not change its direction of variation. As the entropy monotonically depends on the state purity, its decrease
(or increase) does not depend on this noise subtraction,
although its exact value will indeed be changed a bit. We can estimate how
much entropy decreases for our system if we do not do this noise subtraction.
Let us take the third experiment as an example as it has the largest
contribution by noise. With the background noise subtraction, we have
observed an initial entropy about $0.88$ and a final entropy
about $0.76$, which corresponds to an initial Bloch vector
length $L_{0}=0.40$ and the final $L_{f}=0.56$
. Now without background noise subtraction, we should have {%
$L_{0}^{\prime }=p_{0}L_{0}$ and
$L_{f}^{\prime }=p_{0}L_{f}$ by taking $p_{0}$ with its typical value $%
0.84$. The corresponding entropies for the initial and the final states are
given respectively by $0.92$ and $0.83$, so we still have an
observed entropy decrease about $0.09$ bit for this
experiment by taking into account the contribution of the imperfect initial
state polarization.